\documentclass[letterpaper, 10 pt, conference]{ieeeconf}  

\IEEEoverridecommandlockouts                              
\usepackage{graphicx}
\usepackage{amsmath}					
\usepackage{cases}						
\usepackage{amssymb}					
\usepackage{mathtools}					
\usepackage{color}						
\usepackage{bm}							
\usepackage{dsfont}						
\usepackage[american]{circuitikz}		

\usepackage{enumerate}					
\usepackage[utf8]{inputenc}
\usepackage[T1]{fontenc} 
\usepackage{lmodern}					
\usepackage{siunitx}					

\usepackage[caption=false]{subfig}
\usepackage{cancel}

\usepackage{pgfplots}
\pgfplotsset{compat=newest}
\usetikzlibrary{plotmarks}
\usetikzlibrary{arrows.meta}
\usepgfplotslibrary{patchplots}
\usepackage{grffile}

\newtheorem{proposition}{Proposition}
\newtheorem{assumption}{Assumption}
\newtheorem{remark}{Remark}

\usetikzlibrary{shapes,arrows}
\usetikzlibrary{calc}
\usetikzlibrary{angles}
\usetikzlibrary{positioning}
\usetikzlibrary{fit}
\usetikzlibrary{backgrounds}
\usepackage{arydshln}					

	\definecolor{purpleDark}{RGB}{118, 4, 205}
	\definecolor{purpleLight}{RGB}{186, 102, 250}
	
	\definecolor{blueDark}{RGB}{52, 78, 243}
	\definecolor{blueLight}{RGB}{118, 135, 244}
	\definecolor{redDark}{RGB}{197, 34, 0}
	\definecolor{redLight}{RGB}{255, 91, 57}
	\definecolor{yellowDark}{RGB}{255, 183, 0}
	\definecolor{yellowLight}{RGB}{255, 204, 77}
	\definecolor{greenDark}{RGB}{0, 143, 53}
	\definecolor{greenLight}{RGB}{42, 189, 97}

	\colorlet{greenFaint}{green!10!white}
	\colorlet{redFaint}{red!10!white}

	\definecolor{redText}{RGB}{222, 2, 10}
	\definecolor{orangeText}{RGB}{245, 86, 0}
	\definecolor{greenText}{RGB}{20,125,50}
	\definecolor{blueText}{RGB}{0, 114, 190}
	\definecolor{purpleText}{RGB}{115, 38, 146}
	\definecolor{pinkText}{RGB}{255, 107, 250}

	\definecolor{MatlabGreen}{rgb}{0,0.6,0}
	\definecolor{MatlabGray}{rgb}{0.5,0.5,0.5}
	\definecolor{MatlabMauve}{rgb}{0.58,0,0.82}
	
	\colorlet{shadecolor}{gray!15}
	
	\colorlet{colorPhs}{blueText}						
	
	\colorlet{colorIFeed}{blueText}						
	\colorlet{colorIForm}{orangeText}					
	\colorlet{colorISupport}{greenText}					
	\colorlet{colorIFeedForm}{purpleText}				
	\colorlet{colorIFormSupport}{purpleText}			
	\colorlet{colorIFeedFromSupport}{purpleText}		

	\colorlet{colorSLyapDirect}{redText!90!white}
	\colorlet{colorSLyapLaSalle}{orangeText!80!white}
	\colorlet{colorSLyapEquation}{pinkText}
	\colorlet{colorSSynchronisation}{blueText}
	\colorlet{colorSSmallSignal}{purpleText}
	\colorlet{colorSDroopAssumed}{greenText}
	\colorlet{colorSOther}{black}
	
	\colorlet{colorDguComp}{greenDark}					
	\colorlet{colorLineComp}{blueDark}					
	\colorlet{colorLoadComp}{redDark}					
	\colorlet{colorLoadLineComp}{purpleDark}			
	
	\colorlet{colorChange}{blueDark}
	\colorlet{colorCompensate}{redDark}

	\definecolor{colorDGU1}{rgb}{0.00000,0.24000,0.72000}
	\definecolor{colorDGU2}{rgb}{0.85000,0.32500,0.09800}
	\definecolor{colorDGU3}{rgb}{0.92900,0.69400,0.12500}
	\definecolor{colorDGU4}{rgb}{0.49400,0.18400,0.55600}
	\definecolor{colorDGU5}{rgb}{0.00000,0.61000,0.52870}
	
	\colorlet{colorPhasorLoadModel}{greenDark}
	\definecolor{colorPhasorHelpVars}{rgb}{0.85000,0.32500,0.09800} 
	\definecolor{colorPhasorDesiredVars}{rgb}{0.00000,0.44700,0.74100} 
	\definecolor{colorPhasorMeasuredVars}{rgb}{0.49400,0.18400,0.55600} 

	\newcommand{\Matlab}{{\rm \sc Matlab}}			
	\newcommand{\Simulink}{{\rm \sc Simulink}}		
	
	\DeclareSIUnit{\pu}{pu}							
	\DeclareSIUnit{\VAR}{\volt\ampere{}R}			

	\renewcommand{\vec}[1]{\ensuremath{\bm{#1}}}		
	\renewcommand{\matrix}[1]{\ensuremath{\bm{#1}}}		

	\renewcommand{\d}[1]{\text{d}#1}
	\renewcommand*\d{\mathop{}\!\mathrm{d}}

	\newcommand{\Reals}{\ensuremath{\mathbb{R}}}		
	
	\newcommand{\transpose}{\ensuremath{\textsf{T}}}	
	
	\newcommand{\x}{\vec{x}}

\newcommand{\hamiltonian}{\ensuremath{H}}

\newcommand{\Rti}{\ensuremath{R_{\text{t}}}}
\newcommand{\Cti}{\ensuremath{C_{\text{t}}}}
\newcommand{\Ctj}{\ensuremath{C_{\text{t}j}}}
\newcommand{\Lti}{\ensuremath{L_{\text{t}}}}
\newcommand{\Iti}{\ensuremath{I_{\text{t}}}}
\newcommand{\Itidot}{\ensuremath{\dot{I}_{\text{t}i}}}
\newcommand{\ILi}{\ensuremath{I_{\text{L}}}}
\newcommand{\INi}{\ensuremath{I_{\text{N}}}}
\newcommand{\Uti}{\ensuremath{V_{\text{t}}}}
\newcommand{\Ui}{\ensuremath{V}}
\newcommand{\Uidot}{\ensuremath{\dot{V}_i}}
\newcommand{\Uj}{\ensuremath{V_j}}

\newcommand{\ddgu}{d}
\newcommand{\zdgu}{z}

\newcommand{\Rtii}{\ensuremath{R_{\text{t}i}}}
\newcommand{\Ctii}{\ensuremath{C_{\text{t}i}}}
\newcommand{\Ltii}{\ensuremath{L_{\text{t}i}}}
\newcommand{\Itii}{\ensuremath{I_{\text{t}i}}}
\newcommand{\ILii}{\ensuremath{I_{\text{L}i}}}
\newcommand{\INii}{\ensuremath{I_{\text{N}i}}}
\newcommand{\Utii}{\ensuremath{V_{\text{t}i}}}
\newcommand{\Uii}{\ensuremath{V_i}}

\newcommand{\ddgui}{d_i}
\newcommand{\zdgui}{z_i}

\newcommand{\Rline}{\ensuremath{R_{ij}}}
\newcommand{\Lline}{\ensuremath{L_{ij}}}
\newcommand{\Cline}{\ensuremath{C_{ij}}}
\newcommand{\Iline}{\ensuremath{I_{ij}}}
\newcommand{\Ilinedot}{\ensuremath{\dot{I}_{ij}}}
\newcommand{\Rmatrixline}{\ensuremath{\matrix{R}_{ij}}}
\newcommand{\Jline}{\ensuremath{\matrix{J}_{ij}}}

\newcommand{\xline}{\ensuremath{x_{ij}}}

\newcommand{\dline}{\ensuremath{\vec{d}_{ij}}}

\newcommand{\zline}{\ensuremath{\vec{z}_{ij}}}

\newcommand{\YLi}{\ensuremath{Y_{\text{L}}}}

\newcommand{\PLi}{\ensuremath{P_{\text{L}}}}

\newcommand{\YLii}{\ensuremath{Y_{\text{L}i}}}
\newcommand{\PLii}{\ensuremath{P_{\text{L}i}}}

\newcommand{\xeq}[1]{\vec{x}_{#1}^*}
\newcommand{\xeqmg}{\vec{x}_\text{MG}^*}
\newcommand{\Uieq}{V^*}
\newcommand{\Itieq}{\Iti^*}

\newcommand{\Uiieq}{V_i^*}
\newcommand{\Itiieq}{\Itii^*}

\newcommand{\Hdi}{\hamiltonian_{\text{c}}}

\newcommand{\Hdxeins}{\hamiltonian_{\text{c}1}(\x_1)}
\newcommand{\Hdxzwei}{\hamiltonian_{\text{c}2}(\x_2)}

\newcommand{\Jdxii}{\matrix{J}_{\text{c}i}(\x_i)}
\newcommand{\Rdxii}{\matrix{R}_{\text{c}i}(\x_i)}
\newcommand{\Rdxiitransposed}{\matrix{R}_{\text{c}i}^\transpose(\x_i)}

\renewcommand{\xi}{\vec{x}}
\newcommand{\xii}{\vec{x}_i}
\newcommand{\ui}{u}
\newcommand{\vi}{v}
\newcommand{\yi}{y}

\newcommand{\xeinsi}{x_{1}}
\newcommand{\xeqeinsi}{x_{1}^*}
\newcommand{\xzweii}{x_{2}}
\newcommand{\xeqzweii}{x_{2}^*}
\newcommand{\xhieq}{x_{\text{h}}^*}
\newcommand{\xhi}{x_{\text{h}}}

\newcommand{\Hi}{\hamiltonian}
\newcommand{\Hxi}{\hamiltonian(\xi)}
\newcommand{\Hxii}{\hamiltonian_i(\xii)}
\newcommand{\Hdxi}{\hamiltonian_{\text{c}}(\xi)}

\newcommand{\Ji}{\matrix{J}}
\newcommand{\Ri}{\matrix{R}}
\newcommand{\Jdi}{\matrix{J}_{\text{c}}}
\newcommand{\Rdi}{\matrix{R}_{\text{c}}}
\newcommand{\Jdxi}{\matrix{J}_{\text{c}}(\xi)}
\newcommand{\Rdxi}{\matrix{R}_{\text{c}}(\xi)}
\newcommand{\reinsi}{r_{1}}
\newcommand{\reinsii}{r_{1,i}}
\newcommand{\rzweii}{r_{2}}

\newcommand{\betaxi}{\beta(\xi)}
\newcommand{\gi}{\matrix{g}}

\newcommand{\zi}{\vec{z}}
\newcommand{\zeinsi}{z_{1}}
\newcommand{\zeinsidot}{\dot{z}_{1}}
\newcommand{\zeinsieq}{z_{1}^*}
\newcommand{\zhi}{z_{\text{h}}}
\newcommand{\zhieq}{z_{\text{h}}^*}
\newcommand{\zei}{z_{\text{e}}}
\newcommand{\kI}{k_{\text{I}}}
\newcommand{\kIi}{k_{\text{I},i}}

\newcommand{\Hdzi}{\hamiltonian_{\text{cz}}}
\newcommand{\Hdzizi}{\hamiltonian_{\text{cz}}(\zi)}
\newcommand{\Hdzizitilde}{\widetilde{\hamiltonian}_{\text{cz}}(\zi)}

\newcommand{\xeqitilde}{\widetilde{\x}^*}
\newcommand{\xeqeinsitilde}{\widetilde{x}_1^*}

\title{\LARGE \bf
	A Scalable Port-Hamiltonian Approach to Plug-and-Play Voltage Stabilization in DC Microgrids}

\author{Felix Strehle, Martin Pfeifer, Albertus Johannes Malan, Stefan Krebs, S{\"o}ren Hohmann
	\thanks{F.\ Strehle, M.\ Pfeifer, A.J.\ Malan, S.\ Krebs, S.\ Hohmann are with the Institute of Control Systems, Karlsruhe Institute of Technology (KIT), 76131, Karlsruhe, Germany
		{\tt\small felix.strehle@kit.edu}}%
}

\begin{document}
	
	\maketitle
	\thispagestyle{empty}
	\pagestyle{empty}
	
	
	\begin{abstract}                
One of the major challenges of voltage stabilization in converter-based DC microgrids are the multiple interacting units displaying intermittent supply behavior. 
In this paper, we address this by a decentralized scalable, plug-and-play voltage controller for \emph{voltage-source converters} (VSCs) at primary level.
In contrast to existing approaches, we follow a systematic and constructive design based on \emph{port-Hamiltonian systems} (PHSs) which does neither require the heuristic proposition of a Lyapunov function nor the computation of auxilliary variables such as time-derivatives. 
By employing the Hamiltonian naturally obtained from the PHS approach as Lyapunov function and using the modularity of passive systems, we provide sufficient conditions under which the designed VSC controllers achieve microgrid-wide asymptotic voltage stability.
\emph{Integral action} (IA), which preserves the passive PHS structure, robustifies the design against unknown disturbances and ensures zero voltage errors in the steady-state. Numerical simulations illustrate the functionality of the proposed voltage controller.
\end{abstract}

	\section{Introduction}
In recent years, converter-based DC microgrids have been identified as a viable option in future electrical energy supply systems \cite{Tucci16}\cite{Cucuzzella19}\cite{Meng17}. They are local entities comprising flexible loads and \emph{distributed generation units} (DGUs) that also include storage devices. These DGUs are commonly connected to the electrical network via controllable \emph{voltage-source converters} (VSCs) and RLC filters \cite{Meng17}\cite{Schiffer16}\cite{Guerrero11}.

Regarding their control, converter-based microgrids provide a manifold of challenges. A central one is the stabilization of the bus voltages \cite{Schiffer16}\cite{Dragicevic16}, which in DC microgrids directly determine the power flows through the network.
%
At present, the voltage stabilization is structured hierarchically with the basic control task being performed by local VSC controllers at primary level \cite{Meng17}\cite{Schiffer16}\cite{Guerrero11}\cite{Dragicevic16}.
A further challenge of integrating a high share of DGUs is their varying availability due to the intermittent nature of most renewable energy sources, which can in worst case lead to plug-in and -out operations \cite{Sadabadi2018}.

In order to cope with multiple interacting DGUs and their intermittent supply behavior, local VSC controllers at primary level are usually implemented in a decentralized manner \cite[Fig.~2]{Meng17}.
Decentralized control approaches only rely on local information and measurements for the corresponding local VSC control design which (i) makes them independent of the overall microgrid size and thus \emph{scalable}; (ii) drastically simplifies the control design by decomposing the microgrid into more manageable, \emph{modular} subsystems; (iii) allows for the addition or removal of DGUs in a \emph{plug-and-play} fashion without requiring changes to any existing local controllers. From an economic perspective, the purely local information requirement is also compatible with high privacy standards in market systems \cite{Tucci18}.

The most popular decentralized control method at primary level is droop-based voltage control and extensions thereof \cite{Guerrero11}\cite{Dragicevic16}. 
However, the voltage steady-state errors due to primary voltage droop control necessitate distributed secondary controllers \cite{Meng17}\cite{Dragicevic16}. Although the resulting overall control simultaneously achieves offset-free voltage stabilization and current sharing (see \cite{Cucuzzella19consens} and references therein), it comes at the cost of some form of communication and limited plug-and-play capability, and thus is outside the scope of this paper.
An alternative approach to decentralized, primary voltage control  is proposed in \cite{Tucci16} and for example extended in \cite{Sadabadi2018} and \cite{Tucci18}. 
It is based on quasi-stationary approximations of line dynamics which are only valid in low-voltage networks. Furthermore, it requires information about the lines connected to each DGU and solving linear matrix inequalities for each primary voltage controller.

Passivity theory and its close link to Lyapunov stability (cf.\ \cite[pp.~40]{Sepulchre97}) is used in \cite{Cucuzzella19}, \cite{Cucuzzella19distributed}, and \cite{Nahata20} to address the plug-and-play primary voltage control problem in DC microgrids without the need for numerical optimization.
However, they (D1) necessitate a heuristic proposition of an appropriate Lyapunov function which in general might be cumbersome and hampers transfer to other applications. Although \cite{Cucuzzella19} and \cite{Cucuzzella19distributed} follow a more structured proposition via a Krasovskii-type Lyapunov function, this comes at the cost of requiring first-time derivatives which must be estimated in practice in finite-time.
Furthermore, (D2) the approach in \cite{Nahata20} a priori specifies a PI controller which is tuned to shape the proposed Lyapunov function and passivate the system accordingly. In case of system modifications, the complete heuristic process has to be redone. As the attainable performance is restricted from the start to a PI structure, the new problem might be infeasible.

In this paper, we circumvent drawbacks (D1) and (D2) by following a systematic and constructive control design based on  port-Hamiltonian theory and \emph{interconnection and damping assignment passivity-based control} (IDA-PBC). We derive sufficient conditions for the asymptotic voltage stability of the whole DC microgrid by using the modularity of passive systems and the Hamiltonian, naturally obtained by the PHS approach, as a Lyapunov function. This then leads to design requirements for the VSC controllers and, similar to \cite{Cucuzzella19}\cite{Cucuzzella19distributed}\cite{Nahata20}, to scalar inequality conditions for the two-tier Z/ZIP loads connected to each DGU.
In contrast to being restricted PI control structure, our control law in general allows any static state feedback plus an IA. The specific controller then follows naturally from the design process and the set requirements (e.g.\ strict passivity, desired equilibrium) for the closed-loop system.
An additional IA accounts for modeling errors and other disturbances that cause unknown network current flows, which would lead to non-zero voltage errors in the steady-state.

In summary, our main contributions comprise the development of a scalable plug-and-play voltage control law for converter-based DC microgrids which (a) asymptotically stabilizes the voltages in the whole DC microgrid, (b) ensures zero voltage errors in the steady-state (also in the presence of unknown disturbances), and (c) allows a general static state feedback plus IA control structure.
	\section{Preliminaries and Basic Procedure} \label{sec:basicprocedurefundamentals}
%
%
\begin{figure}
	\begin{center}
		\includegraphics[width=1.0\linewidth]{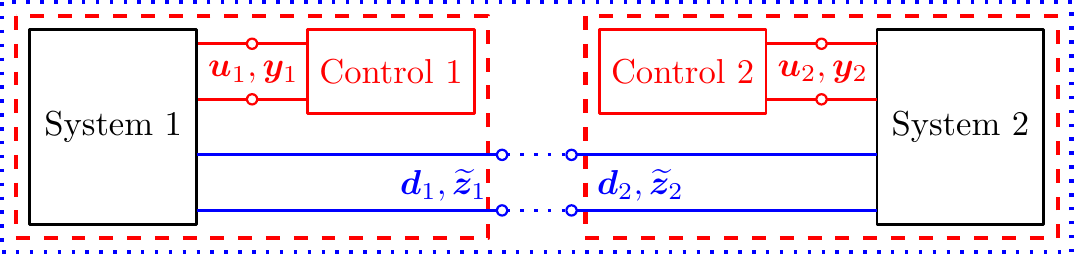}    
		\caption{Interconnection of two (strictly) passive, closed-loop systems (dashed red) via their uncontrollable disturbance ports (blue) to form an autonomous, (strictly) passive and thus (asymptotically) stable overall system (dotted blue)}\label{fig:ModularStabilityPassivity}
	\end{center}
\end{figure}
%

Central to this paper is the interconnection of passive systems modeled as PHSs as illustrated in Fig.~\ref{fig:ModularStabilityPassivity}. 

\subsection{Modular Stability Analysis}
Consider two systems $i\in\{1,2\}$ of port-Hamiltonian form
\begin{subequations}
	\label{eq:phs}
	\begin{align}
	\dot{\x}_i&=\left[ \matrix{J}_i(\x_i)-\matrix{R}_i(\x_i)\right] \frac{\partial H_i(\x_i)}{\partial \x_i}+\matrix{g}_i(\x_i)\vec{u}_i+\matrix{k}_i(\x_i)\vec{d}_i,\\
	\vec{y}_i&=\vec{g}_i^{\textsf{T}}(\x_i)\frac{\partial H_i(\x_i)}{\partial \x_i},\\
	\vec{z}_i&=\vec{k}_i^{\textsf{T}}(\x)\frac{\partial H_i(\x_i)}{\partial \x_i},
	\end{align}
where $\x_i\in\Reals^{n_i}$ is the state vector, $\vec{u}_i \in \Reals^{m_i}$ is the control input vector, $\vec{d}_i \in \Reals^{d_i}$ is the disturbance input vector, $\vec{y}_i\in \Reals^{m_i}$ is the control output vector, and $\vec{z}_i\in \Reals^{d_i}$ is the disturbance output vector \cite[p.~70]{Duindam09}. The matrices $\matrix{J}_i(\x_i)$, $\matrix{R}_i(\x_i)$, $\matrix{g}_i(\x_i)$, and $\matrix{k}_i(\x_i)$ are real-valued matrices of respective sizes with $\matrix{J}_i(\x_i)=-\matrix{J}_i^{\textsf{T}}(\x_i)$ and $\matrix{R}_i(\x_i)=\matrix{R}_i^{\textsf{T}}(\x_i) \succcurlyeq 0$ (positive semidefinite). The Hamiltonian $H_i(\x_i):\Reals^{n_i} \rightarrow \Reals_{\geq 0}$ is a smooth function of the states representing the total energy stored in the system. 
Throughout this paper, we consider quadratic Hamiltonians
\begin{equation} \label{eq:phs:hamiltonian}
H_i(\x_i) = \frac{1}{2}\x_i ^{\textsf{T}} \matrix{Q}_i \x_i,
\end{equation}
\end{subequations}
with $\matrix{Q}_i=\matrix{Q}_i^{\textsf{T}} \succ 0$, i.e.\ \emph{positive definite functions} $H_i(\xeq{i})=0$ and $H_i(\x_i)>0$ for all $\x_i \neq \xeq{i}$ (cf.\ \cite[p.~117]{Khalil02}) with minima specifying the equilibrium of system \eqref{eq:phs} at $\xeq{i} = \arg\min_{\x_i} H_i(\x_i)$.
For $\matrix{R}_i(\x_i) \succ 0$, 
\begin{subequations}
\label{eq:passivityineq}
\begin{align}
\frac{\d \hamiltonian_i(\x_i)}{\d{t}} = \frac{\partial^\transpose \hamiltonian_i(\x_i)}{\partial\x_i} \dot{\x}_i&< \vec{u}_i^\transpose\vec{y}_i+\vec{d}_i^\transpose\vec{z}_i\\
\Leftrightarrow-\frac{\partial^\transpose \hamiltonian_i(\x_i)}{\partial\x_i}  \matrix{R}_i(\x_i) \frac{\partial \hamiltonian_i(\x_i)}{\partial\x_i} & < 0  \quad \forall \x_i \neq \xeq{i}
\end{align}
\end{subequations}
holds and strict passivity of \eqref{eq:phs} w.r.t.\ the supply rate $s_i(\vec{u}_i, \vec{y}_i, \vec{d}_i, \vec{z}_i)=\vec{u}_i^\transpose\vec{y}_i+\vec{d}_i^\transpose\vec{z}_i$ and the Hamiltonian \eqref{eq:phs:hamiltonian} follows \cite[p.~236]{Khalil02}.

Assuming we apply passivity-based control laws to Systems 1 and 2 that shape the energy such that a new Hamiltonian $H_{\text{c}i}(\x_i)$ with new equilibrium $\xeq{i} = \arg\min_{\x_i} H_{\text{c}i}(\x_i)$ is established, and possibly modify the dynamic behavior, we obtain closed-loop PHSs of the form (see dashed red lines in Fig.~\ref{fig:ModularStabilityPassivity}) 
\begin{subequations}
	\label{eq:phs:closedloop}
	\begin{align}
	\dot{\x}_i&=\left[ \Jdxii-\Rdxii\right] \frac{\partial \hamiltonian_{\text{c}i}(\x_i)}{\partial \x_i}+\matrix{k}_i(\x_i)\vec{d}_i,\\
	\widetilde{\vec{z}}_i&=\vec{k}_i^\transpose(\x)\frac{\partial \hamiltonian_{\text{c}i}(\x_i)}{\partial \x_i}
	\end{align}
\end{subequations}
which are again strictly passive w.r.t.\ $s_i(\vec{d}_i, \vec{z}_i)=\vec{d}_i^\transpose\widetilde{\vec{z}}_i$ for positive definite $\hamiltonian_{\text{c}i}(\x_i)$ and  $\Rdxii=\Rdxiitransposed \succ 0$.
If $\vec{d}_i, \widetilde{\vec{z}}_i$ simply specify the interconnection of Systems 1 and 2 via ideal flow (current) or effort (voltage) constraints that algebraically link $\vec{d}_1, \widetilde{\vec{z}}_1$ to $\vec{d}_2, \widetilde{\vec{z}}_2$, the interconnection is a power-preserving Dirac structure \cite[pp.~99]{Duindam09}. In this case, the overall system (see dotted blue in Fig.~\ref{fig:ModularStabilityPassivity}) is again a strictly passive (port-Hamiltonian) system with the equilibrium $\xeq{\text{12}}=\left[\xeq{1}, \xeq{2} \right]^\transpose$. Its Hamiltonian is the sum of the subsystem Hamiltonians
\begin{equation}
\hamiltonian_{\text{12}} (\x_{\text{12}})=\Hdxeins+\Hdxzwei
\end{equation} 
and thus also a positive definite function. As there are no more open ports to interact with the overall system, the supply rate is zero and similar to  \eqref{eq:passivityineq} follows $\forall \x_{12} \neq \x_{12}^*$
\begin{align}
\frac{\d \hamiltonian_{\text{12}}}{\d t}
=-\frac{\partial^\transpose \hamiltonian_{\text{12}}}{\partial \x_{\text{12}}}\left(\matrix{R}_{\text{c}1}(\x_1)+\matrix{R}_{\text{c}2}(\x_2)\right)\frac{\partial\hamiltonian_{\text{12}}}{\partial \x_{\text{12}}}<0.
\end{align}
%
By using this strict passivity w.r.t a zero supply rate and the radially unbounded, positive definite, Hamiltonian $\hamiltonian_{\text{12}}(\x)$ as Lyapunov function, we directly can infer global asymptotic stability of the overall equilibrium $\xeq{\text{12}}$ via Lyapunov's direct method \cite[pp.~44--45]{vdS17}.

\subsection{Basic Procedure}
Based on these considerations, we first model the subsystems \emph{DGU} and \emph{electrical line} comprising any DC microgrid as PHSs of the form \eqref{eq:phs}. Then, we formulate the control problem arising for the requirement of asymptotic voltage stability in a DC microgrid and subsequently design an appropriate decentralized, passivity-based voltage controller.
In a last step, we extended this controller by an IA which preserves the asymptotic voltage stability of the DC microgrid. This robustifies the voltage controller and ensures zero steady-state voltage errors in the presence of model uncertainties and unknown network currents.
For the  remaining part of this contribution, we establish the following assumption which holds under normal grid conditions:
\begin{assumption} \label{assumption:v>0}
	Any voltages (bus, reference, nominal) are strictly positive, i.e.\ $V(t)>0$ for all $t\geq0$. 
\end{assumption}
\begin{remark}
	Note that due to Assumption~\ref{assumption:v>0}, we only use the term \emph{asymptotic voltage stability} in this work and drop the denotation \emph{global}. Although, from a practical perspective, the obtained voltage stability result is global in the sense that it holds for the complete operationally relevant area of $V(t)>0$ for all $t\geq0$.
\end{remark}
\section{Modeling} \label{sec:model}
In Section \ref{subsec:model:system}, we introduce the DC microgrid to be modeled and identify relevant subsystems. Sections \ref{subsec:model:dgu} and \ref{subsec:model:line} present port-Hamiltonian models of the two main subsystems \emph{DGU} and \emph{electrical line}. 

\subsection{System description} \label{subsec:model:system}
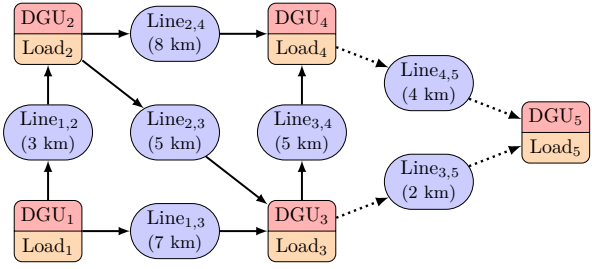
\begin{figure} 
	\centering
	\scalebox{.75}{\begin{tikzpicture}
	\newcommand{\dguTextDGU}[1]{DGU$_{#1}$}
	\newcommand{\dguTextLoad}[1]{Load$_{#1}$}
	\newcommand{\lineText}[3]{Line$_{#1,#2}$ \\ \small (#3 km)}

	\def\dguXdist{4.5cm}
	\def\dguYdist{3.5cm}
	\def\dguNodes{1,2,3,4,5}
	\def\lineNodes{1/2, 1/3, 2/3, 2/4, 3/4, 3/5, 4/5}
	\def\lineNodesLengths{1/2/3, 1/3/7, 2/3/5, 2/4/8, 3/4/5, 3/5/2, 4/5/4}
	
	\def\lineNodesFilled{1/2, 1/3, 2/3, 2/4, 3/4}
	\def\lineNodesDashed{3/5, 4/5}
	
	\tikzstyle{dgu}   = [draw, rectangle split, rounded corners, rectangle split parts=2, rectangle split part fill={red!30, orange!30}]
	\tikzstyle{line}   = [draw, rounded rectangle, minimum height = 0.6cm, fill={blue!20},text width=1cm,align=center]
	\tikzstyle{arrowInFilled}  = [-latex, line width=1.0pt]
	\tikzstyle{arrowOutFilled} = [-latex, line width=1.0pt]
	\tikzstyle{arrowInDashed}    = [-latex, dotted, line width=1.2pt]
	\tikzstyle{arrowOutDashed}   = [-latex, dotted, line width=1.2pt]
	
	\coordinate(cDgu1);
	\path (cDgu1) +(0,\dguYdist) coordinate (cDgu2);
	\path (cDgu1) +(\dguXdist,0) coordinate (cDgu3);
	\path (cDgu2) +(\dguXdist,0) coordinate (cDgu4);
	\path (cDgu3) +(\dguXdist,\dguYdist / 2) coordinate (cDgu5);
	
	\foreach \a/\b in \lineNodes {\coordinate (cLine\a\b) at ($(cDgu\a)!0.5!(cDgu\b)$);}
	
	\foreach \a in \dguNodes {\node[dgu](dgu\a) at(cDgu\a) {\dguTextDGU{\a}\nodepart{second}\dguTextLoad{\a}}; }

	\foreach \a/\b/\l in \lineNodesLengths {\node[line](line\a\b) at(cLine\a\b){\lineText{\a}{\b}{\l}};}
	
	\foreach \a/\b in \lineNodesFilled {
		\draw[arrowInFilled](dgu\a) to (line\a\b);
		\draw[arrowOutFilled] (line\a\b) to (dgu\b);	}
	
	\foreach \a/\b in \lineNodesDashed {
		\draw[arrowInDashed](dgu\a) to (line\a\b);
		\draw[arrowOutDashed] (line\a\b) to (dgu\b);	}
	
%
%
\end{tikzpicture}}
	\caption{
		Bipartite graph representation of a DC microgrid with DGUs, Kron-reduced loads associated with the respective DGUs, and lines interconnecting the various DGUs; dotted lines indicate the plug-and-play nature of the microgrid}
	\label{fig:bipartitegraph}
\end{figure}

We consider a DC microgrid in islanded mode. The microgrid consists of DGUs, loads, and lines in a \emph{load-connected} topology, i.e.\ loads are mapped to the DGU terminals denoted as the \emph{Point of Common Coupling} (PCC) as in \cite{Tucci16}, \cite{Tucci18}, \cite{Cucuzzella19}, \cite{Cucuzzella19distributed} and \cite{Nahata20}. This allows for a bipartite graph representation of the DC microgrid as shown in Fig.~\ref{fig:bipartitegraph} (cf.\ \cite{Nahata20}). The bipartite graph consists of multiple connections of the two subsystems, \emph{DGU} and \emph{line} ($\pi$-model). A single such connection is illustrated in Fig.~\ref{fig:DGUinterface}. Note that each DGU may connect to an arbitrary number of electrical lines. 
%

\subsection{DGU Inverter Interface Model} \label{subsec:model:dgu}
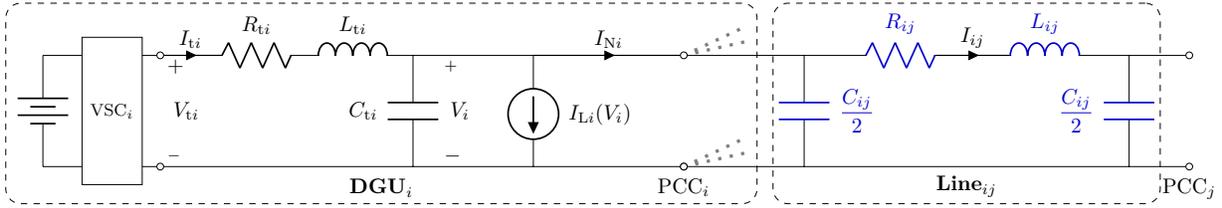
\begin{figure*}
	\centering
	\scalebox{.8}{

\begin{tikzpicture}
	\def\cHeight{1.85cm}	
	\def\colorLine{blue!80!black}	
	\coordinate(dgu_filter_base);
	\draw
		(dgu_filter_base) to [open, o-, v<=$\Utii$] ++(0,\cHeight) coordinate(dgu_filter_base_high)
		to [short, o- ,i=$\Itii$] ++(1.0,0)
		to [R, l=$\Rtii$] ++(1.2,0) 
		to [L, l=$\Ltii$] ++(2.0,0) coordinate(dgu_filter_output_high) 
		to [C, l_=$\Ctii$] ++(0,-\cHeight) coordinate(dgu_filter_output_low) 
		to [short] (dgu_filter_base)
		
		(dgu_filter_output_high) to [open] ++(0.4,0)
		to [open, v^>=$\Uii$] ++(0,-\cHeight)
		
		(dgu_filter_output_high) to  ++(2,0) 	coordinate(dgu_load_high)
		to [short] ++(0,-0.1cm)
		to [american current source, i>^=$\ILii(\Uii)$] ++(0,{-\cHeight+0.1cm}) coordinate(dgu_load_low)
		to [short] (dgu_filter_output_low)
		
		(dgu_load_high) to [short, i^>=$\INii$] ++(2.5, 0) coordinate(dgu_pcc_high)
		to [open] ++(0, -\cHeight) coordinate(dgu_pcc_low)
		to [short, o-] (dgu_load_low)
		
		(dgu_filter_base) to [short] ++(-1.95,0) coordinate(dgu_source_low)
		to [battery, invert] ++(0,\cHeight) coordinate(dgu_source_high)
		to [short] (dgu_filter_base_high);
	
	\path (dgu_filter_base) +(-0.8,\cHeight/2) coordinate (inverter);
	\node[align=center,draw,fill=white,rectangle,minimum height =\cHeight + 0.6cm,minimum width =1.0cm ]() at(inverter) {\footnotesize VSC$_i$};
	
	\node[below=0.08] at (dgu_pcc_low) {PCC$_{i}$};
	
	\draw[loosely dotted, line width=1.55pt, draw=black!50]
		(dgu_pcc_low) to ++(12.5:1.1)
		(dgu_pcc_high) to ++(12.5:1.1)
		
		(dgu_pcc_low) to ++(25:1.1)
		(dgu_pcc_high) to ++(25:1.1);

	\coordinate(line_left_low) at (dgu_pcc_low);
	\coordinate(line_left_high) at (dgu_pcc_high);
	\draw	
		(line_left_high) to [short, o-] ++(2.0,0) coordinate(line_Ci_high)
		to [C, color=\colorLine, \colorLine, l=$\dfrac{\Cline}{2}$] ++(0,-\cHeight) coordinate(line_Ci_low)
		to [short, -o] (line_left_low)
		
		
		(line_Ci_high) to [short] ++(1.0,0)
		to [R, color=\colorLine, \colorLine, l=$\Rline$] ++(1.2,0)
		to [short, i^>=$\Iline$] ++(1.2, 0)
		to [L, color=\colorLine, \colorLine, l=$\Lline$] ++(1.2,0)
		to [short] ++(0.8,0) coordinate(line_Cj_high)
		to [C, color=\colorLine, \colorLine, l_=$\dfrac{\Cline}{2}$] ++(0,-\cHeight) coordinate(line_Cj_low)
		to [short] (line_Ci_low)
		
		(line_Cj_high) to [short, -o] ++(1.0,0) coordinate(line_right_high)
		to [open] ++(0,-\cHeight) coordinate(line_right_low)
		to [short, o-] (line_Cj_low);
		
	\node[below=0.08] at (line_right_low) {PCC$_{j}$};

	\path (dgu_source_low) +(-0.5,-0.5) coordinate (dgu_box_bottom_left);
	\path (dgu_pcc_high) +(1.1,0.75) coordinate (dgu_box_top_right);
	
	\path (line_Ci_low) +(-0.4, -0.5) coordinate (line_box_bottom_left);
	\path (line_Cj_high) +(0.4, 0.75) coordinate (line_box_top_right);
	
	\begin{scope}[on background layer]
		\node[draw,dashed,rounded corners=0.25cm,fit=(dgu_box_bottom_left) (dgu_box_top_right)] (dgu_box) {};
		\node[above] at (dgu_box.south) {\textbf{DGU$_i$}};
		
		\node[draw,dashed,rounded corners=0.25cm,fit=(line_box_bottom_left) (line_box_top_right)] (line_box) {};
		\node[above] at (line_box.south) {\textbf{Line$_{ij}$}};
	\end{scope}
		
\end{tikzpicture}}
	\caption{
		Circuit diagram of a DGU comprising a buck VSC, a filter, and a voltage-dependent current source representing a load, connected to a $\pi$-model line (blue); the legs of the line are considered part of the respective DGUs}
	\label{fig:DGUinterface}
\end{figure*}

The left part of Fig. \ref{fig:DGUinterface} depicts the circuit diagram of a DGU at any node $i$ in the microgrid. It consists of a DC voltage source, which may represent a renewable energy source or a storage device, a buck VSC, and a series RLC filter. The losses in the VSC and filter are lumped together in $\Rtii$. The DC voltage source is assumed to represent an infinite power source \cite{Tucci16}\cite{Tucci18}\cite{Nahata20} and the buck VSC is considered as ideal transformer without operational constraints \cite{Tucci16}\cite{Tucci18}\cite{Nahata20}. The buck VSC is described by an average model using the averaged switch modeling technique \cite{Schiffer16}.
A current source described by $\ILii(\Uii)$ represents the DGU-mapped loads. $\INii$ is the net-current injected into the microgrid  at $PCC_i$ which equals the accumulated incoming and outgoing line currents at $PCC_i$. 
%
%
Based on fundamental electrical network theory, a port-Hamiltonian model of form \eqref{eq:phs} of the DGU VSC interface is defined by
%
\begin{subequations}
	\label{eq:model:DGU:disturbed}
	\begin{align}
		\begin{bmatrix}
		\Ltii\Itidot\\
		\Ctii\Uidot
		\end{bmatrix}&=\begin{bmatrix} 
			-\Rtii 	&	-1\\
			1	&	-\frac{\ILii(\Uii)}{\Uii}
		\end{bmatrix}
		\begin{bmatrix}
		\Itii\\
		\Uii
		\end{bmatrix}+
		\begin{bmatrix}
			1\\
			0
		\end{bmatrix}\Utii-
		\begin{bmatrix}
			0	\\
			1	
		\end{bmatrix}
		\INii, \label{eq:model:dgux}\\ 
		\Itii&=
		\begin{bmatrix}
			1	&	0
		\end{bmatrix}
		\begin{bmatrix}
		\Itii\\
		\Uii
		\end{bmatrix}, \label{eq:DGUyfull}\\ 
		\Uii&=
		\begin{bmatrix}
			0	&	1
		\end{bmatrix}
		\begin{bmatrix}
		\Itii\\
		\Uii
		\end{bmatrix} \label{eq:DGUzfull},\\
		\Hxii&=\frac{x_{1i}^2}{2 \Ltii}+\frac{x_{2i}^2}{2 \Ctii}=\frac{\Ltii}{2}\Itii^2+\frac{\Ctii}{2}\Uii^2,
\label{eq:DGUHamiltonian}
	\end{align}
%
%
with $\xii = \left[\Ltii\Itii,\Ctii\Uii \right]^\transpose$, $u_i=\Utii$, $y_i=\Itii$, $\ddgui =-\INii$, $\zdgui =\Uii$ and interconnection and damping matrices
\begin{equation}\label{eq:model:JRdgu}
\matrix{J}_i=\begin{bmatrix}
0& -1\\
1& 0
\end{bmatrix}, \quad \matrix{R}_i(\x_i)=\begin{bmatrix}
\Rtii & 0\\
0 & \frac{\ILii(\Uii)}{\Uii}
\end{bmatrix}.
\end{equation}
\end{subequations}
Note that from a modeling point of view, $\INii$ is an unknown disturbance. The load $\ILii(\Uii)$ is already integrated into \eqref{eq:model:DGU:disturbed} and injects damping (see \eqref{eq:model:JRdgu}) dependent on its characteristics and the bus voltage $\Uii$. 

\subsubsection{Load Model} In this work, we model the load connected to the $PCC_i$ of a DGU at node $i$ by a two-tier Z/ZIP model (\cite[pp.~110-112]{Machowski2008}). For voltages above $0.7 V_0$ ($V_0$: nominal microgrid voltage), we use the static, polynomial ZIP model. It comprises the parallel combination of a constant impedance (Z) expressed as admittance (Y), a constant current (I), and a constant power (P) load
\begin{equation} \label{eq:model:loadcurrent}
I_{\text{L}i}(V_i)=Y_{\text{L}i} V_i+\bar{I}_{\text{L}i}+P_{\text{L}i} V_i^{-1}, \quad\Uii\geq 0.7 V_0
\end{equation}
with $Y_{\text{L}i}>0$, $\bar{I}_{\text{L}i}=const.>0$, and $P_{\text{L}i}>0$. For voltages below $0.7 V_0$, the corresponding Z load forms a special case of the subsequent analysis with $\bar{I}_{\text{L}i}=0$ and $P_{\text{L}i}=0$ in \eqref{eq:model:loadcurrent}. Consequently, it is not considered explicitly.

\subsection{Electrical line model}  \label{subsec:model:line}
The right side of Fig. \ref{fig:DGUinterface} shows the structure of an exemplary electrical line from node $i$ to $j$. The line is described by a $\pi$-model with parameters $\Rline$, $\Lline$, $\Cline >0$. Shunt conductances are neglected. The line capacitances may be lumped with the respective filter capacitances $\Ctii$ and $\Ctj$. However, even for an exemplary \SI{50}{\kilo\meter} transmission line with the electric constant $\epsilon_0$, ground distance $D=\SI{5}{\meter}$, and radius $r=\SI{0,02}{\meter}$
\begin{equation} \label{eq:model:line:capacitance}
\Cline\approx\frac{2\pi\epsilon_0}{ln(\frac{\SI{5}{\meter}}{\SI{0,02}{\meter}})}\SI{50}{\kilo\meter}=\SI{0,504}{\micro\farad}\ll \SI{2,2}{\milli\farad}\approx C_{\text{t}i/j}
\end{equation}
(cf.\ \cite[p.~201]{Kundur94}\cite[Table 1]{Tucci16}). Thus their influence may be neglected for the control design, ensuring the local buck VSC controllers are independent of these unknown line parameters. From the resulting RL model one can easily obtain a one-dimensional PHS \eqref{eq:phs} with
%
\begin{subequations}
	\label{eq:model:line} 
\begin{align} 
\Lline\Ilinedot&=\left[ -\Rline\right] \left[\Iline \right] +\begin{bmatrix}
1 & -1
\end{bmatrix} \begin{bmatrix}
\Uii\\
\Uj
\end{bmatrix},\\
\begin{bmatrix}
\Iline\\
-\Iline
\end{bmatrix}&=\begin{bmatrix}
1\\
-1
\end{bmatrix} \Iline,\\
H_{ij}(\xline)&=\frac{\Lline}{2}\Iline^2,
\end{align}
\end{subequations}
where $\xline =\Lline\Iline$, $\Jline = 0$, $\Rmatrixline = \Rline$, $\dline = \left[\Uii, \Uj\right]^\transpose $, and $\zline = \left[\Iline, -\Iline\right]^\transpose$. Note that \eqref{eq:model:line} has no controlled ports. 
\section{Problem Formulation} \label{subsec:model:connection}
%
The considered control problem is finding plug-and-play control laws for the VSCs of the DGUs which asymptotically stabilize the bus voltages $\Uii$, i.e.\ $\lim\limits_{t\to \infty}\Uii = \Uiieq$ for all $i=1,\dots,N$. The  references $\Uiieq$ are specified by a higher-level control. With \eqref{eq:model:DGU:disturbed}, this implies the desired closed-loop DGU equilibria
\begin{equation} \label{eq:model:equilibriumdgu}
\xeq{i}=\left[\Lti\Itiieq, \Cti\Uiieq \right]^\transpose, \quad i=1,\dots,N.
\end{equation}
%
The current references $x_{1,i}^*=\Ltii\Itiieq$ are not specified explicitly and follow as a consequence of the load demand and network exchange currents (see second row of \eqref{eq:model:dgux}). As outlined in Section~\ref{sec:basicprocedurefundamentals}, the strict passivity and thus asymptotic stability of the equilibrium of a system can be inferred by analyzing the passivity of the subsystems and their interconnections.
\begin{proposition}\label{prop:mgvoltagestability}
	A DC microgrid represented by a bipartite graph as in Fig.~\ref{fig:bipartitegraph} comprising $RL$-lines and strictly passive DGU PHSs is itself strictly passive with an asymptotically stable equilibrium $\xeqmg$ given by the combined equilibria of the subsystems \emph{DGU} (including its connected load) $\xeq{i}$ and \emph{electrical line} $\xeq{ij}$, respectively.
\end{proposition}
\proof
According to Section \ref{sec:basicprocedurefundamentals}, the line PHS in \eqref{eq:model:line} is a strictly passive system w.r.t.\ its ports $\dline, \zline$ as $\Rmatrixline=\Rline>0$ per definition. If the DGU model \eqref{eq:model:DGU:disturbed} is strictly passive w.r.t.\ its respective ports $\ddgui,\zdgui$ (possibly through appropriate control to establish the desired equilibria \eqref{eq:model:equilibriumdgu} and modify the dynamic behavior), then the DC microgrid comprises only strictly passive subsystems (cf.\ Fig.~\ref{fig:ModularStabilityPassivity} dashed red) interconnected by ideal flow (current) constraints which are power-conserving Dirac structures (\cite[p.~100]{Duindam09}). The strict passivity of the DC microgrid then directly follows from \cite[p.~107]{Duindam09}. Its asymptotic voltage stability can be inferred from Lyapunov's direct method (\cite[p.~44]{vdS17}) by using the Hamiltonian of the DC microgrid $\hamiltonian_{\text{MG}}$, i.e.\ the sum of the subsystem Hamiltonians, as a Lyapunov function. As the strictly passive DC microgrid has no more open ports (cf.\ Fig.~\ref{fig:ModularStabilityPassivity} dotted blue), $\dot{\hamiltonian}_{\text{MG}}<0, \forall\x_{\text{MG}}\neq\xeqmg$ holds. 
\QEDclosed

From Proposition~\ref{prop:mgvoltagestability} follows that the task of voltage stabilization in DC microgrids reduces to two main problems: (P1) Appropriately controlling the local buck VSCs such that their respective DGUs are strictly passive w.r.t. $\ddgui,\zdgui$ and the minima of their Hamiltonians are at the desired voltage references $\Uiieq=x_{2,i}^*/\Ctii , \; i=1,\dots,N$. This also includes setting up inequalities for the control parameters and load characteristics to guarantee the strict passivity; (P2) Ensuring zero steady-state errors of the bus voltages under model uncertainties and other unmodeled disturbances.	
\section{Plug-and-play voltage control design}
In Sections~\ref{sec:control:pbc} and \ref{sec:control:closedloopstability}, we address problem (P1) by first designing a passivity-based voltage controller for the initially undisturbed DGU model, i.e.\ \eqref{eq:model:DGU:disturbed} with $\ddgui=0$. Then we establish inequalities for the control parameters and load characteristics to ensure strict passivity of the \emph{DGU} subsystem and thus microgrid-wide asymptotic voltage stability as per Proposition~\ref{prop:mgvoltagestability}. Problem (P2) is addressed in Section~\ref{sec:control:ia}, where we extend the passivity-based voltage controller with an IA that preserves the closed-loop PHS form and thus the asymptotic voltage stability. 
%
\begin{assumption} \label{assumption:piecewiseconstant}
	In the sequel, we assume voltage references $\Uiieq$ and disturbances $\ddgui$ to be piecewise constant w.r.t.\ the fast VSC dynamics such that suitable constant equilibria exist.
\end{assumption}
\begin{remark}
For clarity in the following control design, the index $i$ referring to the $i$-th DGU is dropped from all variables and parameters, i.e.\ $\Ui=\Uii$, $\Rti=\Rtii$, etc.  
\end{remark}

%
%
\subsection{Passivity-Based Voltage Controller} \label{sec:control:pbc}
%
%
\begin{proposition} \label{prop:IDA}
Consider a DGU model as defined in \eqref{eq:model:DGU:disturbed}. Suppose the disturbance currents are zero for the control design, i.e.\ $\ddgu=-\INi=0$. If \eqref{eq:model:DGU:disturbed} is controlled by the static state feedback control law
	%
\begin{subequations}
	\label{eq:ida}
	\begin{align}	
	\ui(\x)=\betaxi&=\frac{\left(\Rti-\reinsi \right)}{\Lti}\xeinsi+\frac{\xeqzweii}{\Cti}+\reinsi\ILi\left(\frac{\xeqzweii}{\Cti}\right), \\
	&=\left(\Rti-\reinsi \right)\Iti+\Uieq+\reinsi\ILi(\Uieq). \label{eq:IDAcontrollaw_voltages}
	\end{align}
\end{subequations}
	then its closed-loop dynamics are given by
	\begin{subequations}
		\label{eq:closedloopphs}	
		\begin{align} 
		\begin{bmatrix}
		\Lti\Itidot\\
		\Cti\Uidot
		\end{bmatrix}&=\begin{bmatrix} 
		-\reinsi	&	-1\\
		1	&	-\rzweii(\Ui)
		\end{bmatrix}
		\begin{bmatrix}
		\Iti-\Itieq\\
		\Ui-\Uieq
		\end{bmatrix}-
		\begin{bmatrix}
		0	\\
		1	
		\end{bmatrix}
		\INi,\\
		\Ui-\Uieq&=
		\begin{bmatrix}
		0	&	1
		\end{bmatrix}
		\begin{bmatrix}
		\Iti-\Itieq\\
		\Ui-\Uieq
		\end{bmatrix}
		\end{align}
	\end{subequations}
	with the closed-loop equilibrium
	\begin{equation} \label{eq:equilibrium}
	\xeq{}=\left[\Lti\Itieq, \Cti\Uieq \right]^\transpose.
	\end{equation}
	The constant damping $\reinsi\geq0$ is the control parameter, $\Uieq=\xeqzweii/\Cti$ is the voltage reference, and the state $\xeinsi=\Lti\Iti$ or rather its respective current $\Iti$ is a measurement.
%
%
\end{proposition}
%

\proof
In order to obtain \eqref{eq:ida}, we apply an IDA-PBC approach \cite{Ortega04}. In general, IDA-PBC design assigns a desired closed-loop PHS of the form
\begin{subequations}
	\label{eq:desiredphs}
	\begin{align}
	\xi&=\left[ \Jdxi-\Rdxi\right] \frac{\partial \Hdxi}{\partial \xi}\\
	\yi&=\gi^\transpose \frac{\partial \Hdxi}{\partial \xi}
	\end{align}
\end{subequations}
with $\xeq{} = \arg\min_{\xi} \Hdxi$ by means of a static state feedback
\begin{equation}
\ui(\xi)=\left[\gi^\transpose\gi\right]^{-1}\!\!\gi^\transpose\!
\left( \left[ \Jdi-\Rdi\right] \!\frac{\partial \Hdi}{\partial \xi} - \left[ \Ji-\Ri\right] \!\frac{\Hi}{\partial \xi}\right) \label{eq:ida:generallaw}
\end{equation}
This control law is obtained after solving the IDA-PBC matching equation
\begin{equation}
\begin{split}
\gi^{\perp}(\xi)\left[ \Jdxi-\Rdxi\right] \frac{\partial\Hdxi}{\partial \xi}=\\
\gi^{\perp}(\xi)\left[ \Ji(\xi)-\Ri(\xi)\right] \frac{\partial \Hxi}{\partial\xi}.
\end{split} \label{eq:ME}
\end{equation}
Equation~\eqref{eq:ME} represents a system of linear, first-order PDEs whose solutions specify the assignable $\Hdxi$ for fixed $\Jdxi$ and $\Rdxi$. However, by fixing the desired $\Hdxi$, \eqref{eq:ME} can be simplified and becomes an algebraic system of equations in $\Jdi$, $\Rdi$, $\gi^{\perp}$. Accordingly, we fix the desired Hamiltonian to
\begin{equation}
\Hdxi =\frac{\Lti}{2}(\Iti-\Itieq)^2+\frac{\Cti}{2}\left(\Ui-\Uieq \right)^2,
\label{eq:Hamiltoniandesired}
\end{equation}
by performing the simplest possible shift to \eqref{eq:DGUHamiltonian} such that $\xeq{} = \arg\min_{\xi} \Hdxi$ is established at \eqref{eq:equilibrium}.
For the solution of \eqref{eq:ME}, we structurally parameterize the closed-loop interconnection and damping matrices with
\begin{equation} \label{eq:JRparameterization}
\Jdi=\begin{bmatrix}
0				&	-J_{12}\\
J_{12}	&	0
\end{bmatrix},
\Rdi=
\begin{bmatrix}
\reinsi	&	0\\
0			& \rzweii(\xi)	
\end{bmatrix}
\end{equation}
where $\reinsi$, $\rzweii(\xi)\geq0$ $\forall\, \xi$.
Selecting the natural choice for the full-rank left annihilator $\gi^{\perp}=\left[0, 1 \right]$ since $\matrix{g}_i=\left[1, 0 \right]$, and inserting \eqref{eq:model:DGU:disturbed} with $\ddgu=0$, \eqref{eq:Hamiltoniandesired}, \eqref{eq:JRparameterization}, and $\gi^{\perp}$ into the matching equation \eqref{eq:ME} yields
\begin{align}
\begin{bmatrix}
J_{12} &\! -\rzweii(\xi)
\end{bmatrix}\!\!
\begin{bmatrix}
\Iti-\Itieq\\
\Ui-\Uieq
\end{bmatrix}
&=\begin{bmatrix}
1 & \!-\frac{\ILi(\Ui)}{\Ui}
\end{bmatrix}\!\!
\begin{bmatrix}
\Iti\\
\Ui
\end{bmatrix}
,\\
J_{12}(\Iti-\Itieq)-\rzweii(\xi)(\Ui-\Uieq)&=\Iti-\ILi(\Ui). \label{eq:MEinserted}
\end{align}
%
%
Then we rewrite the load current equation \eqref{eq:model:loadcurrent} as
\begin{align} 
\ILi(\Ui)-\ILi(\Uieq)=\YLi(\Ui-\Uieq)+\PLi\left(\frac{1}{\Ui}-\frac{1}{\Uieq}\right)\\
 \Leftrightarrow \ILi(\Ui)=
\left[\YLi-\frac{\PLi}{\Ui\Uieq} \right] (\Ui-\Uieq)+\ILi(\Uieq)
\label{eq:rewriteload}
\end{align}
and insert \eqref{eq:rewriteload} in \eqref{eq:MEinserted}. By comparing the coefficients for $(\Iti-\Itieq)$ and $(\Ui-\Uieq)$ in \eqref{eq:MEinserted}, we obtain
\begin{align}
J_{12}(\Iti-\Itieq)&=\Iti-\ILi(\Uieq)\\
\rzweii(\xi)(\Ui-\Uieq)&=
\left[\YLi-\frac{\PLi}{\Ui\Uieq} \right] (\Ui-\Uieq)
\end{align}
from which $\Itieq=\ILi(\Uieq)$, $J_{12}=1$, and
\begin{equation}\label{eq:control:r2}
\rzweii(\xi)=\rzweii(\Ui)=
\YLi-\frac{\PLi}{\Ui\Uieq}, \quad\Ui, \Uieq\geq 0.7 V_0
\end{equation}
follow. Finally, with \eqref{eq:model:DGU:disturbed} for $\ddgu=0$, \eqref{eq:Hamiltoniandesired}, and \eqref{eq:JRparameterization}, the static state feedback controller \eqref{eq:ida} can be calculated. 
%
%
The closed-loop dynamics \eqref{eq:closedloopphs} of the DGU model \eqref{eq:model:DGU:disturbed} are obtained by inserting \eqref{eq:IDAcontrollaw_voltages} in \eqref{eq:model:DGU:disturbed}. \QEDclosed
%
\begin{remark}\label{remark:idainterpretation}
Note that the control law \eqref{eq:ida} comprises a damping assignment via static state feedback $\left(\Rti-\reinsi \right)\Iti$ to establish a new filter resistance $\reinsi$, and a model-based disturbance compensation of the load influence in steady-state $\reinsi\ILi(\Uieq)$. 
\end{remark}

\subsection{Closed-Loop Voltage Stability} \label{sec:control:closedloopstability}
According to Proposition~\ref{prop:mgvoltagestability}, the asymptotic voltage stability of the DC Microgrid is established if the closed-loop \emph{DGU} subsystem \eqref{eq:closedloopphs} is strictly passive w.r.t.\ its port $\ddgu,\zdgu$. As outlined in Section~\ref{sec:basicprocedurefundamentals}, strict passivity can be inferred from $\matrix{R}(\x)\succ0$. With
\begin{equation} \label{eq:Rdinequalities}
\Rdxi=\begin{bmatrix} 
\reinsi	&	0\\
0	&	\rzweii(\Ui)
\end{bmatrix}\overset{!}{\succ}0,
\end{equation}
from \eqref{eq:closedloopphs}, we obtain with \eqref{eq:control:r2} the inequalities
\begin{align}
\reinsi&>0, \label{eq:ineq_control}\\
\rzweii(\Ui)&>0 \Leftrightarrow
\YLi\Ui\Uieq>\PLi, &\Ui, \Uieq\geq 0.7 V_0.
 \label{eq:ineq_load}
\end{align}
The control parameter $\reinsi\geq0$ represents a degree of freedom in the design, which allows to adjust the damping of the $\xeinsi$-dynamics. It can thus be chosen such that \eqref{eq:ineq_control} is always fulfilled. 
However, for the ZIP load model in \eqref{eq:ineq_load}, the instantaneous voltage $\Ui$ and the voltage reference $\Uieq$ must not drop below a lower boundary specified by the load characteristics $\YLi, \PLi$ in order to ensure strict passivity. This consequently restricts the operating area at each $PCC_i$ depending on the characteristics of the connected load. As \eqref{eq:ineq_load} is most restrictive at $\Ui,\Uieq=0.7 V_0$, we propose
\begin{equation}
	0.49\YLi V_0^2>\PLi \label{eq:ineq_load_conservative}
\end{equation}
as a conservative measure to ensure the strict passivity of \eqref{eq:closedloopphs} for any $\Ui, \Uieq \in \left[0.7 V_0, \infty\right)$ under given $\YLi$, $\PLi$. In light of Proposition~\ref{prop:mgvoltagestability}, \eqref{eq:ineq_load_conservative} is thus a sufficient condition for the asymptotic voltage stability obtained under control law \eqref{eq:ida}.


\subsection{Addition of Integral Action} \label{sec:control:ia}
\begin{figure}
\centering
\scalebox{0.6}{\tikzset{
block/.style = {draw, fill=white, rectangle, minimum height=3em, minimum width=3em},
tmp/.style  = {coordinate}, 
sum/.style= {draw, fill=white, circle, node distance=1cm},
input/.style = {coordinate},
output/.style= {coordinate},
pinstyle/.style = {pin edge={to-,thin,black}
}
}

\begin{tikzpicture}[auto, node distance=1.5cm,>=latex', font=\Large]
\node [input, name=rinput] (rinput) {};
\node [sum, right of=rinput] (sum0) {};
\node [block, above of=sum0,node distance=1.3cm] (r1loadcomp){$\reinsi\ILi(\Uieq)$};
\node [sum, right of=sum0, node distance=1.5cm] (sum1) {};
\node [block, right of=sum1] (gain_L) {\LARGE $\frac{1}{\Lti}$};
\node [block, right of=gain_L, node distance=2cm] (integrator_L) {$\int$};
\node [block, above of=gain_L,node distance=1.3cm] (rti){$\reinsi$};
\node [sum, right of=integrator_L,node distance=2cm] (sum2) {};
\node [sum, right of=sum2, node distance=1.2cm] (sum3) {};
\node [block, above of=sum3,node distance=1.3cm] (load){$\ILi(\Ui)$};
\node [block, right of=sum3] (gain_C) {\LARGE $\frac{1}{\Cti}$};
\node [block, right of=gain_C, node distance=2cm] (integrator_C) {$\int$};
\node [output, right of=integrator_C,] (output) {};
\node [tmp, below of=integrator_L,node distance=1.3cm] (tmp1){};
\node [tmp, right of=integrator_L,node distance=0.8cm] (tmp2){};
\draw [->] ($(rinput)+(+0.5,0)$) -- node[left]{$\Uieq$} (sum0);
\draw [->] (r1loadcomp) --(sum0);
\draw [->] (sum0) --(sum1);
\draw [->] (sum1) --(gain_L);
\draw [->] (gain_L) -- node{$\dot{I}_{\text{t}}$} (integrator_L);
\draw [->] (integrator_L) -- node{$\Iti$} (sum2);
\draw [->] (tmp2) |-  (rti);
\draw [->] (sum2) --(sum3);
\draw [->] (load) --node[pos=0.90] {$-$} (sum3);
\draw [->] (sum3) -- (gain_C);
\draw [->] (gain_C) -- node{$\dot{V}$}(integrator_C);
\draw [->] (integrator_C) -- node [name=y] {$\Ui$}(output);
\draw [->] (rti) -| (sum1);
\draw [dotted,->] (y) |-(load);
\draw [->] (y) |-(tmp1)-| node[pos=0.95] {$-$} (sum1);
\draw [->] ($(0,1.5cm)+(sum2)$)node[above of=sum2, node distance=2cm]{$\ddgu$} -- (sum2);
\end{tikzpicture}}
\caption{Block diagram of the closed-loop DGU model \eqref{eq:closedloopphs}}
\label{fig:blockdiagIDA}
\end{figure}
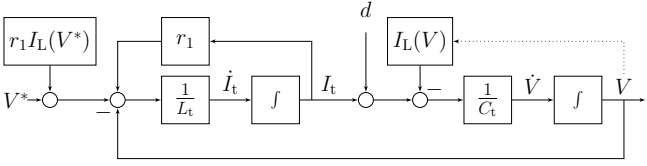

As outlined in Remark~\ref{remark:idainterpretation}, the control law \eqref{eq:ida} uses the load model \eqref{eq:model:loadcurrent} to compensate the steady-state load influence. Naturally, exact model knowledge is neither fully accurate nor fully available. Furthermore, the damping assignment of $\reinsi>0$ (see \eqref{eq:ineq_control}) which ensures strict passivity and thus asymptotic voltage stability, introduces a zero steady-state voltage error under non-vanishing disturbances $\ddgu=-\INi \neq 0$. This fact can be understood by considering the block diagram of the closed-loop DGU model \eqref{eq:closedloopphs} given in Fig.~\ref{fig:blockdiagIDA}. 
Neglecting the load influence $\ILi(\Ui)$, we obtain the linear disturbance transfer function
\begin{equation}
\frac{\Ui(s)}{\ddgu(s)}=\frac{\Lti s +\reinsi}{\Lti\Cti s^2+\reinsi\Cti s + 1}
\end{equation}
For piecewise constant disturbances $\ddgu(s)=\frac{\ddgu}{s}$ follows
\begin{equation}
	\lim_{s\to 0+} s\frac{\ddgu}{s}\frac{\Ui(s)}{\ddgu(s)}=\reinsi\ddgu.
\end{equation}
Thus, an additional IA is necessary to robustify the control and to guarantee zero-steady state errors of the bus voltage. For this purpose, the control \eqref{eq:ida} is extended by IA via $\vi$ to
\begin{equation} \label{eq:controllaw_general}
\ui(\xi)=\Uti=\betaxi+\vi
\end{equation}
(cf.\ \cite[p.~445]{Ortega04}) yielding the closed-loop DGU dynamics
%
\begin{align} \label{eq:ia:closeddgudynamics}
\begin{bmatrix}
\Lti\Itidot\\
\Cti\Uidot
\end{bmatrix}&=\begin{bmatrix} 
-\reinsi	&	-1\\
1	&	-\rzweii(\Ui)
\end{bmatrix}
\begin{bmatrix}
\Iti-\Itieq\\
\Ui-\Uieq
\end{bmatrix}-
\begin{bmatrix}
0	\\
1	
\end{bmatrix}
\INi+
\begin{bmatrix}
1\\
0
\end{bmatrix}\vi
\end{align}
with \eqref{eq:control:r2}.
Since the disturbances act via $\ddgu=-\INi$ on $\dot{x}_{2}=\Cti\Uidot$, which is not directly actuated by $\vi$, the standard integral feedback of the passive output as suggested in \cite{Ortega04} is not applicable and the IA via state transformation from \cite{Donaire09} is employed.
%
\begin{proposition}
	Consider the closed-loop DGU dynamics \eqref{eq:ia:closeddgudynamics} obtained with the static state feedback \eqref{eq:ida}. Suppose the disturbance current $\ddgu=-\INi$ is piecewise constant (see Assumption~\ref{assumption:piecewiseconstant}). Then, the IA
	\begin{subequations}
	\label{eq:IAcontrollaw}
	\begin{align} 
	\vi&=\frac{\kI\reinsi}{\Cti}\int \left( \xeqzweii - \xzweii \right) \d{t} + \frac{\kI\Lti}{\Cti} \left( \xeqzweii - \xzweii\right)\\
	&=\kI\reinsi\int (\Uieq-\Ui)\d{t}+\kI\Lti(\Uieq-\Ui),
	\end{align}
	\end{subequations}
	establishes the new equilibrium
	\begin{equation} \label{eq:ia:equilibrium}
	\xeqitilde=\left[\xeqeinsitilde, \xeqzweii \right] ^\transpose=\left[\Lti(\ILi(\Uieq)+\INi), \Cti\Uieq \right]^\transpose.
	\end{equation}
	Note that this IA does not affect the voltage stability of the closed-loop DGU model \eqref{eq:closedloopphs} as it preserves $\xeqzweii=\Cti\Uieq$ and the port-Hamiltonian structure assigned by IDA-PBC, i.e.\ $\Jdi$, $\Rdxi$, and the shaped $\Hdxi$. Only the entries $\kI$ in the interconnection matrix and a quadratic term of the extended state $\zei$ in the Hamiltonian are added (cf.\ \eqref{eq:ia:zsystem}).
\end{proposition}
%

\proof
For $\ddgu=-\INi$, \eqref{eq:ia:closeddgudynamics} is in its canonical form (cf.\ \cite[(2)]{Donaire09}\cite[(12)]{Kotyczka13}) with one ($n_1=1$) \emph{actuated} or \emph{relative-degree-one (RD1)} state $\xeinsi$, which is directly actuated by the controlled input (here $\vi$), and one \emph{higher relative-degree (HRD)} or \emph{unactuated} state $x_h=\xzweii$ that receives no direct action through $\vi$. 

Initially, we extend \eqref{eq:ia:closeddgudynamics} with one additional integrator state
\begin{equation} \label{eq:integral}
\zei=\kI \int \frac{\partial \Hdzizi}{\partial \zhi} \d{t}=\kI \frac{1}{\Cti} \int \left(\zhi-\zhieq\right) \d{t}.
\end{equation}
and rewrite it in new $z$-coordinates as
\begin{align}
\begin{bmatrix}
\dot{z}_{1}\\
\dot{z}_{\text{h}}\\
\dot{z}_{\text{e}}
\end{bmatrix}&=\begin{bmatrix} 
-\reinsi	&	-1	&	0\\
1			&	-\rzweii(\zhi)	&	-\kI\\
0			&	\kI	&	0	
\end{bmatrix}
\begin{bmatrix} 
\partial \Hdzizi/\partial \zeinsi\\
\partial \Hdzizi/\partial \zhi\\
\partial \Hdzizi/\partial \zei
\end{bmatrix} \nonumber\\
\Hdzizi&=\Hdi(\zeinsi, \zhi)+\frac{1}{2k}\zei^2 \label{eq:ia:zsystem}\\
&=\frac{1}{2\Lti}(\zeinsi-\zeinsieq)^2+\frac{1}{2\Cti}\left(\zhi-\zhieq \right)^2+\frac{1}{2k}\zei^2. \nonumber
\end{align}

Afterwards, we establish the HRD state transformation
\begin{equation} \label{eq:HRDtrafo}
\xhi=\zhi
\end{equation}
such that the equilibrium in terms of $\zhieq$ matches the desired one $\xhieq$ implying $\Uieq$. Subsequently, we find the RD1 state transformation which satisfies requirement \eqref{eq:HRDtrafo} by solving
\begin{align}
\dot{x}_{\text{h}}&=\dot{z}_{\text{h}} \nonumber\\
\frac{\partial \Hdi}{\partial \xeinsi}-\rzweii(\xzweii)\frac{\partial \Hdi}{\partial \xzweii}&=\frac{\partial \Hdzi}{\partial \zeinsi}-\rzweii(\zhi)\frac{\partial \Hdzi}{\partial \zhi}-\kI\frac{\partial \Hdzi}{\partial \zei} \nonumber\\
\frac{\xeinsi-\xeqeinsi}{\Lti}&=\frac{\zeinsi-\zeinsieq}{\Lti}-\frac{\kI}{k}\zei
\end{align}
for $\zeinsi\eqqcolon \Psi(\xeinsi, \xhi,\zei)$ which yields
\begin{equation} \label{eq:RD1trafo}
\zeinsi=\xeinsi-\xeqeinsi+\zeinsieq+\frac{\kI\Lti}{k}\zei .
\end{equation}

Finally, we compute the integral control law $\vi$ with \eqref{eq:ia:closeddgudynamics} from
\begin{align}
\zeinsidot&\overset{!}{=}\frac{\Psi(\xeinsi, \xhi,\zei)}{\d{t}}\\
-\reinsi \frac{\zeinsi-\zeinsieq}{\Lti}-\frac{\zhi-\zhieq}{\Cti}
&=-\reinsi\frac{\xeinsi-\xeqeinsi}{\Lti}-\frac{\xzweii-\xeqzweii}{\Cti} +\vi \nonumber\\ 
&\phantom{=}+\frac{\kI^2\Lti}{\Cti k}\left( \zhi-\zhieq\right)  \label{eq:pbc:keliminate}
\end{align}
by inserting \eqref{eq:integral}, \eqref{eq:HRDtrafo}, \eqref{eq:RD1trafo}, $x_\text{h}=\xzweii$, and solving for $\vi$. By setting $k=\kI$ in \eqref{eq:pbc:keliminate} as it offers no more degrees of freedom here, we obtain the IA \eqref{eq:IAcontrollaw}.
The new equilibrium that is established for \eqref{eq:ia:closeddgudynamics} under the acting constant disturbance $\ddgu=-\INi$ with the IA \eqref{eq:IAcontrollaw} is investigated by rewriting \eqref{eq:ia:zsystem} as
\begin{align}
\begin{bmatrix}
\dot{z}_{1}\\
\dot{z}_{\text{h}}\\
\dot{z}_{\text{e}}
\end{bmatrix}&=\begin{bmatrix} 
-\reinsi	&	-1	&	0\\
1			&	-\rzweii(\zhi)	&	-\kI\\
0			&	\kI	&	0	
\end{bmatrix}
\begin{bmatrix} 
\partial \Hdzizitilde/\partial \zeinsi\\
\partial \Hdzizitilde/\partial \zhi\\
\partial \Hdzizitilde/\partial \zei
\end{bmatrix} \nonumber\\
\Hdzizitilde&=\frac{1}{2\Lti}(\zeinsi-\zeinsieq)^2+\frac{1}{2\Cti}\left(\zhi-\zhieq \right)^2 \label{eq:ia:zsystem_disturbed}\\
&\phantom{=}+\frac{1}{2\kI}(\zei-\ddgu)^2. \nonumber
\end{align}
From \eqref{eq:HRDtrafo} and \eqref{eq:ia:zsystem_disturbed} follows that $\zhieq=\xhieq=\Cti\Uieq$ and $\zei^*=\ddgu=-\INi$. In order for $\zeinsi\rightarrow\zeinsieq$ to hold, it follows from \eqref{eq:RD1trafo} that
\begin{equation}
	\xeinsi\rightarrow\xeqeinsi-\Lti\zei^*=\Lti\left(\ILi(\Uieq)+\INi\right)\eqqcolon \xeqeinsitilde,
\end{equation}
which yields \eqref{eq:ia:equilibrium}. \QEDclosed

The overall control input \eqref{eq:controllaw_general} defining $\Uti$ is now calculated with \eqref{eq:IDAcontrollaw_voltages} and \eqref{eq:IAcontrollaw} to
\begin{subequations}
\label{eq:controllaw_overall} 
\begin{align} 
\ui&=\frac{(\Rti-\reinsi)}{\Lti}\xeinsi+\frac{\xeqzweii}{\Cti}+\reinsi\ILi\left(\frac{\xeqzweii}{\Cti} \right) \nonumber\\
&\phantom{=}+\frac{\kI\reinsi}{\Cti}\int (\xeqzweii-\xzweii)\d{t}+\frac{\kI\Lti}{\Cti}(\xeqzweii-\xzweii),\\
&=(\Rti-\reinsi)\Iti+\Uieq+\reinsi\ILi(\Uieq) \nonumber\\
&\phantom{=}+\kI\reinsi\int (\Uieq-\Ui)\d{t}+\kI\Lti(\Uieq-\Ui).
\end{align}
\end{subequations}
%
It can be seen that \eqref{eq:controllaw_overall} amounts to a static state feedback w.r.t\ $\xeinsi=\Lti\Iti$, a model-based compensation of the steady-state load influence $\reinsi\ILi(\Uieq)$ (see Remark~\ref{remark:idainterpretation}), and a linear PI-controller w.r.t.\ the HRD state $\xzweii=\Cti\Ui$, which represents the IA. The control parameters are $\reinsi>0$ and $\kI>0$.
\begin{remark}
Note that in absence of any load model knowledge, the steady-state load compensation $\reinsi\ILi(\Uieq)$ is simply omitted. Instead, the PI-controller compensates for it at the cost of impaired performance.
\end{remark}
\section{Simulation}
%
In this section, we assess the functionality of the local passivity-based voltage controller with IA \eqref{eq:controllaw_overall} and its performance by a simulation with \emph{SimScape} in \Matlab/\Simulink.
For this, we use the DC microgrid in Fig.~\ref{fig:bipartitegraph} consisting of five DGUs connected by seven lines. Each DGU comprises a VSC, an RLC filter, and a load \eqref{eq:model:loadcurrent} (cf.\ Fig.~\ref{fig:DGUinterface}).
All DGU filters are parameterized identically with $\Rti=\SI{0.2}{\ohm}$, $\Lti=\SI{1.8}{\milli\henry}$, $\Cti=\SI{2.2}{\milli\farad}$ for $i=1,\dots,5$ (cf.\ \cite{Tucci16}).
In order to investigate the robustness of the control design, full $\pi$-model lines are used with the default \Matlab \ per-kilometer values $\Rline=\SI{0.01273}{\ohm \per \km}$, $\Lline=\SI{0.9337}{\milli\henry \per \km}$, $\Cline=\SI{12.74}{\nano\farad \per \km}$ and lengths as given in Fig.~\ref{fig:bipartitegraph}.
The control parameters are chosen to $\reinsii=5\Rtii$ and $\kIi = \SI{500}{\second^{-1}}$. The former allows an intuitive interpretation of the damping assignment as it is tantamount to replacing the original filter resistances with a five times higher value.
Reference voltages are set around $V_0=\SI{50}{V}$ (cf.\ \cite{Tucci16}) and the load parameters are chosen such that they satisfy \eqref{eq:ineq_load_conservative} (cf.\ Table~\ref{table:simulation_dgu_params}). 
%
%
\begin{table}[!t]
	\centering
	\renewcommand{\arraystretch}{1.25}
	\caption{
		Reference voltages and load parameters of the DGUs (parameters in brackets indicate values after $t = \SI{3}{\second}$)}
	\label{table:simulation_dgu_params}
	\begin{tabular}{@{\,\,}l@{\;\!}c@{\!\quad\quad}c@{\,}}
		\noalign{\hrule height 1pt}
		& & \\[-10pt]
		DGU & $\Uiieq(\si{\volt})$ & $\YLii (\si{\ohm^{-1}}),\ILii(\si{\ampere}),\PLii(\si{\watt})$ \\
		\hline
		1 \textbf{\color{colorDGU1}(blue)} & 50 & 1/2, 1, 200  \\
		2 \textbf{\color{colorDGU2}(red)} & 49.8 & 1/6, 1, 80  \\
		3 \textbf{\color{colorDGU3}(yellow)} & 49.9 & 1/8, 1, 100   \\
		4 \textbf{\color{colorDGU4}(purple)} & 49.7 & 1/10, 1, 50 (1/10, 1, 100)  \\
		5 \textbf{\color{colorDGU5}(turquoise)} & 50.1 & 1/4, 1, 150 \\
		\noalign{\hrule height 1.0pt}
	\end{tabular}
\end{table}
The simulation starts off with DGU~5 disconnected from the microgrid. To test the feasibility of a plug-and-play operation, DGU~5 is connected at $ t = \SI{2}{\second} $ as indicated in Fig.~\ref{fig:bipartitegraph}. To test the robustness of the microgrid, an additional $\SI{50}{\watt}$ load is connected to DGU 4 (cf.\ Table~\ref{table:simulation_dgu_params}) at $t = \SI{3}{\second}$.
%
\begin{figure}
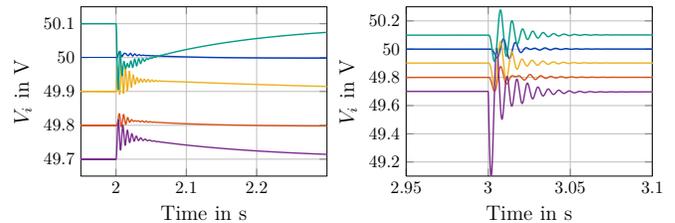

	\centering
	\subfloat {\scalebox{.66}{\input{img/simulation_results_voltages_DGU5connect.tex}}}
	\subfloat {\scalebox{.66}{\input{img/simulation_results_voltages_DGU4loadjump.tex}}}
	\caption{Simulation results of the $\Uii$ for DGU 5 connection at $t=\SI{2}{\second}$ and load step at DGU 4 at $t=\SI{3}{\second}$; DGU colors are given in Table~\ref{table:simulation_dgu_params}}
	\label{fig:simulation_results}
\end{figure}
The resulting bus voltages in Fig.~\ref{fig:simulation_results} indicate stability and zero-steady state errors in both cases. For the plug-in of DGU~5 at $t = \SI{2}{\second}$, the largest deviation with approx.\ $\SI{-0.2}{\volt}$ occurs at DGU~5. But with approx.\ $\SI{-0.4}{\percent}$, it is well within a $\pm\SI{10}{\percent}$ band around its reference. The constant power load step at DGU~4 at $t = \SI{3}{\second}$ induces larger, oscillating voltage deviations, as the load damping in the DC microgrid is reduced. With approx.\ $\SI{-0.6}{\volt}= \SI{-1.2}{\percent}$, however, they are still well within the $\pm\SI{10}{\percent}$ band and decay fast in less than $\SI{50}{\milli\second}$.
	\section{Conclusion}

In this paper, we presented a new systematic and constructive control design for scalable, plug-and-play voltage stabilization in DC microgrids. Based on port-Hamiltonian modeling and the modularity of passive systems, we provide sufficient conditions for the VSC control design and the loads that lead to microgrid-wide asymptotic voltage stability.  
As we use the Hamiltonian as natural Lyapunov function, we avoid the possibly cumbersome heuristic proposition of a Lyapunov function that is mandatory in existing approaches. 
Furthermore, the IDA-PBC design allows a general static state feedback control structure which we enable with an additional IA to handle unknown disturbances and model uncertainties, if the resulting current flows can be considered as piecewise constant over time. 
Future work will address an extension to current flows that are not piecewise constant as well as operational constraints on the power converters.
	
	

\end{document}